# A conditionally integrable bi-confluent Heun potential involving inverse square root and centrifugal barrier terms

## T.A. Ishkhanyan[1,2], V.P. Krainov[2], and A.M. Ishkhanyan[1,3]

[1]Institute for Physical Research, NAS of Armenia, 0203 Ashtarak, Armenia
[2]Moscow Institute of Physics and Technology, Dolgoprudny, 141700 Russia
[3]Physic and Technology Institute, National Research University Tomsk Polytechnic University, Tomsk, 634050 Russia

We present a conditionally integrable potential, belonging to the bi-confluent Heun class, for which the Schrödinger equation is solved in terms of the confluent hypergeometric functions. The potential involves an attractive inverse square root term $\sim x^{-1/2}$ with arbitrary strength and a repulsive centrifugal barrier core $\sim x^{-2}$ with the strength fixed to a constant. This is a potential well defined on the half-axis. Each of the fundamental solutions composing the general solution of the Schrödinger equation is written as an irreducible linear combination, with non-constant coefficients, of two confluent hypergeometric functions. We present the explicit solution in terms of the non-integer order Hermite functions of scaled and shifted argument and discuss the bound states supported by the potential. We derive the exact equation for the energy spectrum and approximate that by a highly accurate transcendental equation involving trigonometric functions. Finally, we construct an accurate approximation for the bound-state energy levels.

## 1. Introduction

We present a bi-confluent Heun potential for which the Schrödinger equation is solved in terms of the Hermite functions which are functions of the confluent hypergeometric class. The potentials allowing the solution of wave equations in terms of the hypergeometric functions have been successfully applied to explore the non-classical evolution during the whole course of the development of quantum mechanics starting already from the seminal papers by Schrödinger [1-8]. However, it has recently been shown that the set of currently known potentials that are exactly or conditionally integrable in terms of these functions (the term conditionally integrable refers to the cases when the potential parameters are not varied independently, see, e.g., [9-14]) almost exhausts the list of all possible potentials for which the solution is achieved through a one-term ansatz involving a single hypergeometric function [15].

A notable progress is then achieved if one appeals to the five functions of the Heun class [16-18] which are direct generalizations of the hypergeometric functions. The treatment in terms of the Heun functions suggests a large set of new potentials (see, e.g., [15,19-25]) which in several cases present generalizations of the classical hypergeometric potentials or their conditionally integrable extensions, while in other cases describe potential families that do not posses hypergeometric sub-potentials.

Though the mathematical theory of the Heun functions is currently underdeveloped, however, notably, in several cases these functions allow reductions to linear combinations of the familiar hypergeometric functions (see, e.g., [26-32]). One then arrives at a new set of hypergeometric potentials the peculiarity of which is that the solution of the Schrödinger equation is no more written through the simple one-term ansatz widely used in the past. Rather, the general solution now involves independent fundamental solutions each of which is an irreducible linear combination of two or more hypergeometric functions. In this way, the list of the exactly solvable hypergeometric potentials is currently enlarged to involve the third ordinary hypergeometric [33], inverse square root [34] and Lambert-W step-barrier [35] potentials that we have introduced recently. In addition, many new conditionally integrable hypergeometric potentials have been also proposed [34-39].

In the present paper we introduce one more conditionally integrable potential. The solution of the Schrödinger equation for this potential employs the Hermite functions which belong to the confluent hypergeometric class. The potential involves the inverse square root potential $V_1 / \sqrt{x}$ with arbitrary strength $V_1$ and a centrifugal-barrier term $V_{cf} / x^2$ with the strength being fixed to $V_{cf} = 21\hbar^2 / (32m)$. The potential belongs to the bi-confluent Heun class with $m_1 = -1$ (see [15,30]). The solution for this potential is derived by expanding the solution of the bi-confluent Heun equation in terms of the Hermite functions and terminating the series on the fifth term [30]. The resultant solution is thus written as a linear combination with constant coefficients of five non-integer order Hermite functions of a scaled and shifted argument. Using the recurrence relations satisfied by the involved Hermite functions, the combination is further reduced to a two-term one with non-constant coefficients. We present the explicit solution and derive the exact spectrum equation for the bound states supported by the potential. Applying a specific asymptotic expansion for the involved Hermite functions, we approximate the spectrum equation by an accurate transcendental equation involving trigonometric functions and construct a highly accurate approximation for the bound-state energy levels applicable to the whole variation range of the parameters. The indexes of the Hermite functions involved in the bound-state wave functions are not integers so that the wave functions are not polynomials or quasi-polynomials.

## 2. The potential and the general solution

The one-dimensional stationary Schrödinger equation for a particle of mass $m$ and energy $E$ is written as

$$\frac{d^2\psi}{dx^2} + \frac{2m}{\hbar^2}\big(E - V(x)\big)\psi = 0,$$ (1)

where $\hbar$ is the reduced Planck constant, and the potential $V(x)$ we treat is

$$V = V_0 + \frac{V_1}{\sqrt{x}} + \frac{21\hbar^2}{32m}\frac{1}{x^2}.$$ (2)

The shape of the potential is shown in Fig. 1. Owing to the inverse square root term, this is a long-range potential which for negative $V_1 < 0$ presents a well and, hence, supports bound states. Since the integral $I_B = \int_0^\infty x\,|V(x)|\,dx$ diverges, the number of bound states is infinite [40,41]. A reason for the potential to be of interest is that owing to the involved centrifugal-barrier term it models, to a certain extent, the one-dimensional reduction of the three-dimensional Schrödinger problem for the central inverse square root potential $1/\sqrt{r}$. However, since the strength of the centrifugal-barrier term is fixed to a specific value, this is a conditionally integrable potential. The potential is a member of a bi-confluent Heun family [22] (that with $m_1 = -1$, see [15]). Apart from the inverse square root potential, which is an exactly integrable one [34], this family possesses several known conditionally integrable confluent hypergeometric sub-potentials, among which are the second Stillinger potential [9], Eq. (2.20), and the López-Ortega potential [37] with its generalization involving the square root potential [39]. The latter one is equivalent to the first potential by Exton [42], Eq. (21).

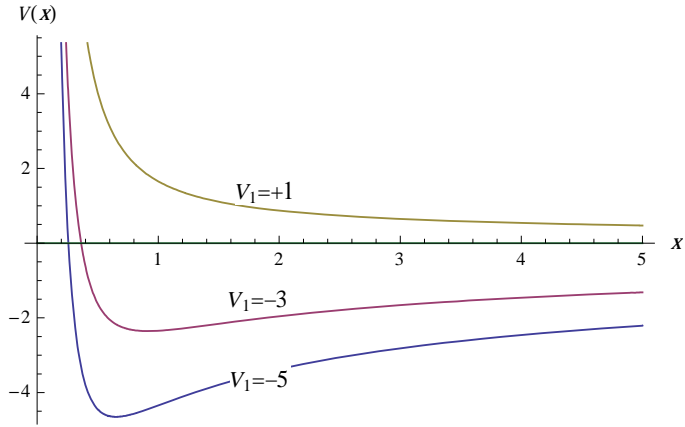

Fig.1 Potential (2) versus $V_1$ ($m = \hbar = 1$, $V_0 = 0$).

The potential and the corresponding solution are derived by reducing the Schrödinger equation to the bi-confluent Heun equation [16-18] as described in [15,22], and further expanding the solution of the latter equation in terms of the Hermite functions as described in [30]. Requiring the series to terminate on the fifth term and allowing the potential parameters to be dependent, one arrives at two potentials one of which (the one which is a sub-potential of the bi-confluent family with $m_1 = -1$ [15,22]) is just potential (2). The solution of the Schrödinger equation for this potential is a linear combination of five contiguous Hermite functions, which is reduced, using the recurrence relation $H_{\nu+1}(z) = 2zH_\nu(z) - 2\nu H_{\nu-1}(z)$, to a combination involving just two Hermite functions. Because of the $z$-proportional term of the recurrence relation, the resultant combination is with non-constant coefficients. The details of the derivation of the potential as well as the solution of the Schrödinger equation are presented below in section **4**.

The result of derivations is that a fundamental solution of the Schrödinger equation for real negative $V_1 < 0$ and $E < V_0$ is written as

$$\psi_F(x) = x^{-3/4} e^{-\left(\sqrt{2a} - \sqrt{\varepsilon x}\right)^2/2} u(x),\tag{3}$$

with

$$u(x) = \left(\sqrt{2a} + \sqrt{\varepsilon x}\right) H_{a+1/2}\left(\sqrt{2a} - \sqrt{\varepsilon x}\right) - (1+2a)(1-\varepsilon x) H_{a-1/2}\left(\sqrt{2a} - \sqrt{\varepsilon x}\right),\tag{4}$$

where $H_a$ is the Hermite function and the involved parameters are given as

$$\varepsilon = \sqrt{\frac{8m(-E+V_0)}{\hbar^2}},\tag{5}$$

$$a = \frac{m^2 V_1^2 / \hbar}{\left(2m(-E+V_0)\right)^{3/2}}.\tag{6}$$

It is further checked that a second independent solution is constructed using equations (3),(4) by applying the change $\varepsilon \to -\varepsilon$, $a \to -a$. Hence, the general solution of the Schrödinger equation is written as the linear combination

$$\psi(x) = C_1 \psi_F + C_2 \psi_F\big|_{\varepsilon \to -\varepsilon, a \to -a}\tag{7}$$

with arbitrary constant coefficients $C_{1,2}$. We note that the solution can alternatively be written in terms of the confluent hypergeometric functions.

### 3. Bound states

The bound states are derived by demanding the wave function to vanish in the origin and at the infinity. The first condition leads to the relation

$$C_2 = -e^{-2a} \frac{\sqrt{2a}\, H_{a+1/2}\left(\sqrt{2a}\right) - H_{a+3/2}\left(\sqrt{2a}\right)}{\sqrt{-2a}\, H_{-a+1/2}\left(\sqrt{-2a}\right) - H_{-a+3/2}\left(\sqrt{-2a}\right)} C_1, \qquad (8)$$

while the second one results in the following equation for the parameter $a$:

$$\left(1 + 2a\right) H_{a-1/2}\left(-\sqrt{2a}\right) + \sqrt{2a}\, H_{a+1/2}\left(-\sqrt{2a}\right) = 0. \qquad (9)$$

This equation possesses a countable infinite set of discrete positive roots $a_n > 0$ defining the energy spectrum through equation (6). We note that, since $H_0 = 1$ and $H_1 = 2z$, $a = 1/2$ is a root of this equation. However, this root does not produce a bound state because it leads to an identically zero wave function. Indeed, for $a = 1/2$ the first independent fundamental solution $\psi_F(x)$ given by equations (3),(4) identically vanishes everywhere and it follows from equation (7) that $C_2 = 0$, so that we get $\psi(x) \equiv 0$. The next root is close to $3/2$. It produces a bound state with one extremum. Hence, this corresponds to the ground state to which we prescribe the number $n = 1$.

To construct an appropriate approximation for the spectrum equation (9), following the lines of the treatment for the inverse square root potential [34], we divide the equation by $(1 + 2a) H_{a-1/2}\left(-\sqrt{2a}\right)$ and apply the large-argument series expansions for the involved Hermite functions. In doing this, one should be careful that the indexes and the arguments of both Hermite functions $H_\nu(z)$ involved in equation (9) belong to the *left* transient layer for which $z \approx -\sqrt{2\nu + 1}$ [43]. For this layer, the argument of the Hermite function is negative so that the expansions of standard reference books are not applicable (see, e.g., [43]). This is understood if one appeals to the following representation of the Hermite function through the confluent hypergeometric functions:

$$H_\nu(z) = \frac{2^{\nu+1}\sqrt{\pi}}{\Gamma(-\nu/2)}\left(\left(\sqrt{z^2} - z\right)\,_1F_1\left(\frac{1-\nu}{2};\frac{3}{2};z^2\right) + \frac{\Gamma(-\nu/2)}{2\sqrt{\pi}}U\left(-\frac{\nu}{2};\frac{1}{2};z^2\right)\right), \qquad (10)$$

where $_1F_1$ and $U$ are the Kummer and Tricomi functions, respectively. We note that for a real positive $z > 0$ the first term of this representation vanishes, while for $z < 0$ it does not. The standard expansions assume the Hermite functions being polynomials and treat the left transient layer by application of the identity $H_n(-z) = (-1)^n H_n(z)$, $n \in \mathbb{N}$. However, in our case the Hermite functions are not polynomials (the polynomial reductions, which are achieved for half-integer $a$, do not produce physically acceptable bound sates because they do not vanish in the origin). An appropriate approximation applicable for the left transient layer is presented in [44]. We here apply an extension of this approximation written as

$$H_\nu(z) \approx h(\nu, z) \left( \cos(\pi \nu) \operatorname{Ai}\left( t - \frac{B_0}{\nu} \right) - \sin(\pi \nu) \operatorname{Bi}\left( t + \frac{B_0}{\nu} \right) \right), \tag{11}$$

where Ai and Bi are the Airy functions and

$$h(\nu, z) = e^{\frac{z^2}{2} - \frac{A_0}{\nu}} 2^{\frac{\nu}{2} + \frac{1}{4}} \frac{1}{\pi^{\frac{1}{4}}} \nu^{-\frac{1}{12}} \sqrt{\Gamma(\nu + 1)}, \tag{12}$$

$$t = -2^{1/2} \nu^{1/6} \left( z + \sqrt{2\nu + 1} \right). \tag{13}$$

Compared with [44], we have modified the arguments of the involved Airy functions and the exponential term involved in the pre-factor $h(\nu, z)$ by adding the correction terms proportional to the parameters $A_0$ and $B_0$ which are supposed to be small. These parameters are to be varied in order to better meet the deviation from the strict transient point $z = -\sqrt{2\nu + 1}$ for the particular values of $\nu$ and $z$ of the Hermite functions involved in equation (9). This variation is achieved by matching the exact and approximate values at a point close to the origin, say, at $a = 3/2$ for simplicity of calculations. Though the needed variation is small, however, it is rather useful in order to achieve an approximation uniformly applicable to the whole permissible variation range of the parameter $a$. Since for the first Hermite function holds $z = -\sqrt{2\nu + 1}$, it is understood that in this case no variation is needed so that we take $A_0 = B_0 = 0$. It is further checked that for the second Hermite function the appropriate choice is $A_0, B_0 \approx 1/5$ (see below for the improved value of $B_0$).

Using further the expansions for the Airy functions in the vicinity of the origin (for large values of parameter $a$ the arguments of the Airy functions are equal to or are close to zero), for the involved Hermite functions we have

$$H_{a-1/2}\left(-\sqrt{2a}\right) \approx 2h_1 \sin(\pi a + \pi/3), \tag{14}$$

$$H_{a+1/2}\left(-\sqrt{2a}\right) \approx -2h_2 \left( \frac{\sin(\pi a - \pi/3)}{6B_0 \sqrt[3]{a}} + \sin(\pi a + \pi/3) \right), \tag{15}$$

with

$$B_0 = \frac{\Gamma(1/3)}{6\sqrt[3]{3}\Gamma(2/3)} \approx \frac{1}{5}. \tag{16}$$

Now we divide the spectrum equation (9) by $(1+2a)H_{a-1/2}\left(-\sqrt{2a}\right)$ and apply equation (12) to calculate the ratio $h_2/h_1$. As a result, we get the approximation

$$F \equiv 1 + \frac{\sqrt{2a}H_{a+1/2}\left(-\sqrt{2a}\right)}{(1+2a)H_{a-1/2}\left(-\sqrt{2a}\right)} \approx f(a)\left(\frac{3B_0}{a^{2/3}} - \frac{\sin\left(\pi a - \pi/3\right)}{\sin\left(\pi a + \pi/3\right)}\right) \qquad (17)$$

with a pre-factor function $f(a)$ which does not adopt zeros apart from the rudimentary root $a=1/2$. To a good extent, this function is approximated as $f(a) \approx a^{-4/3}(2a-1)/(12B_0)$. With the accurate $f(a)$ and $B_0 = 1/5$, equation (17) provides a fairly good approximation for the whole variation range $a \geq 1/2$. The accuracy of the result is demonstrated in Fig. 2, where the filled circles present the exact numerical values and the solid curves stand for the derived approximation.

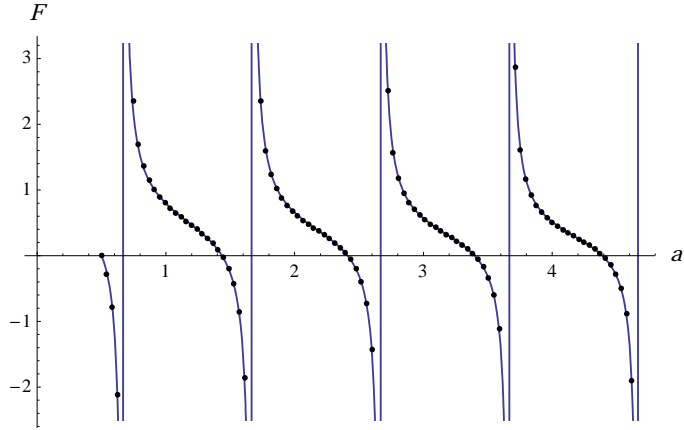

Fig. 2. Filled circles - exact numerical result, solid curves - approximation (17).

Since the pre-factor $f(a)$ does not adopt zeros for $a > 1/2$, we conclude that the spectrum equation is well approximated by the transcendental equation

$$\frac{3B_0}{a^{2/3}} - \frac{\sin\left(\pi a - \pi/3\right)}{\sin\left(\pi a + \pi/3\right)} = 0. \qquad (18)$$

To treat this equation, we note that the first term is always small. Neglecting the term for large $a$, we get the zero-order approximation as $a_n \approx n+1/3$. This result stands for the semiclassical limit, the Maslov index being equal to 1/3 [44,45] (note that for potential (2) the effective angular momentum such that $l(l+1)=21/16$ is $l=3/4$). We note that $a_n$ are not integers, hence, the bound-state wave-functions are not polynomials.

An accurate approximation for the roots of equation (18) can be constructed by replacing $a$ in the first term by $n + 1/3$ and resolving the equation with respect to $a$. The result reads

$$a = -\frac{1}{\pi}\tan^{-1}\left(\sqrt{3}\,\frac{\nu+1}{\nu-1}\right), \tag{19}$$

where

$$\nu = \frac{3B_0}{\left(n+1/3\right)^{2/3}}, \tag{20}$$

Expanding the $\tan^{-1}$ function in equation (19) in terms of powers of $\nu$ and keeping the first two terms, we arrive at a rather accurate approximation (see the resultant accuracy for the energy spectrum in the inset of Fig. 3):

$$a_n \approx n + \frac{1}{3} + \frac{b_1}{\left(n+1/3\right)^{2/3}} - \frac{b_2}{\left(n+1/3\right)^{4/3}}, \quad n = 1,2,3,... \tag{21}$$

with

$$b_1 = \frac{3\sqrt{3}}{10\pi} \approx \frac{1}{6}, \quad b_2 = \frac{3\sqrt{3}}{100\pi} \approx \frac{1}{20}. \tag{22}$$

The relative error of this approximation is less than $10^{-4}$ for all orders $n \geq 1$, the absolute error reaching $2 \times 10^{-4}$ only for the second root with $n = 2$.

According to equation (6), the energy spectrum is given as (we put $V_0 = 0$)

$$E_n = -\left(\frac{mV_1^4}{8\hbar^2}\right)^{1/3}\frac{1}{a_n^{2/3}}, \quad n = 1,2,3,.... \tag{23}$$

With the expansion (21), this provides energy levels with relative error less than $2 \times 10^{-4}$. The corresponding first three normalized wave functions for the parameters $V_1 = 1$, $V_0 = 0$ with $m = \hbar = 1$ are shown in Fig. 4.

A concluding remark is as follows. It is understood that $E_n$ permits expansion in powers of $\left(n+1/3\right)^{-2/3}$. Using expansion (21), it is revealed that this expansion has the form

$$E_n \approx -\left(\frac{mV_1^4}{8\hbar^2}\right)^{1/3}\left(\frac{1}{\left(n+1/3\right)^{2/3}} + \frac{d_1}{\left(n+1/3\right)^{7/3}} + \frac{d_2}{\left(n+1/3\right)^3} + ...\right). \tag{24}$$

Adjusting now $d_1 = -1/(5\sqrt{3})$ and $d_2 = 1/(13\sqrt{3})$ provides relative error less than $10^{-5}$ for all energy levels (see the inset of Fig. 3). Comparing the derived energy spectrum with that for the bare inverse square root potential, we see that the main difference is that here the Maslov semi-classical correction index is $+1/3$ while it is $-1/6$ for the inverse square root potential [34]. This is of course because of the centrifugal term involved in potential (2). Finally, it should be noted that for the first ten levels expansion (24) is by three orders of magnitude more accurate than the semiclassical result $E_n \sim \left(n + 1/3\right)^{-2/3}$.

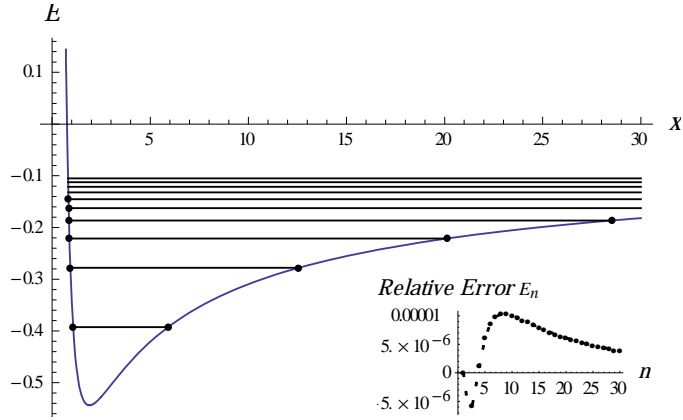

Fig. 3. The energy spectrum of potential (2) for $V_1 = -1$ ( $m = \hbar = 1$, $V_0 = 0$ ).

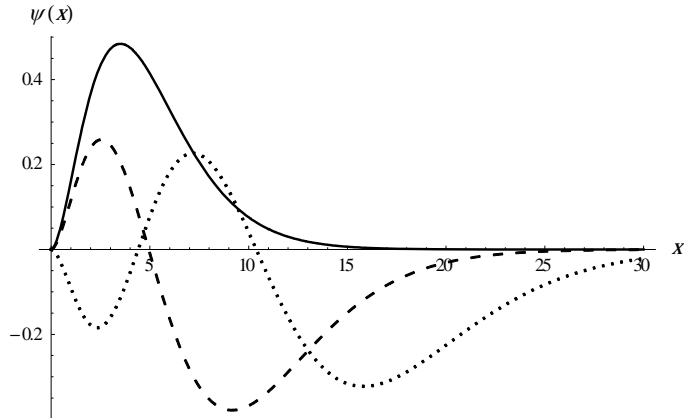

Fig. 4. The first three normalized wave functions for $V_1 = 1$ ( $m = \hbar = 1$, $V_0 = 0$ ).

## 4. Derivation of the potential and the first fundamental solution

For completeness of the treatment, we here present the derivation of potential (2) and the construction of the fundamental solution (3)-(6).

As already mentioned above, the potential under consideration is a member of the first bi-confluent Heun family presented by Lemieux and Bose [22] (see also [15,30]). In its general form this family involves four inverse-power terms with arbitrary strengths:

$$V(x) = V_0 + \frac{V_1}{\sqrt{x}} + \frac{V_2}{x} + \frac{V_3}{x^{3/2}} + \frac{V_4}{x^2}.$$
(25)

By application of the transformation

$$\psi = z^{\alpha_0} e^{\alpha_1 z + \alpha_2 z^2} u(z) \tag{26}$$

with

$$z = \sqrt{2x} , \tag{27}$$

the Schrödinger equation (1) for potential (25) is reduced to the bi-confluent Heun equation [16-18]

$$\frac{d^2 u}{dz^2} + \left( \frac{\gamma}{z} + \delta + \varepsilon z \right) \frac{du}{dz} + \frac{\alpha z - q}{z} u = 0 \tag{28}$$

with the involved parameters given as [30]

$$\gamma = 2\alpha_0 - 1, \ \delta = 2\alpha_1, \ \varepsilon = 4\alpha_2 , \tag{29}$$

$$\alpha = \alpha_1^2 + 2(\gamma + 1)\alpha_2 - \frac{4mV_2}{\hbar^2}, \quad q = -\gamma\alpha_1 + \frac{4\sqrt{2}mV_3}{\hbar^2} \tag{30}$$

and

$$\alpha_0^2 = 2\alpha_0 + \frac{8mV_4}{\hbar^2}, \quad \alpha_1 = \frac{mV_1}{\sqrt{2}\,\hbar^2\alpha_2}, \quad \alpha_2^2 = \frac{m(V_0 - E)}{2\hbar^2} . \tag{31}$$

To solve the bi-confluent Heun equation (28), we apply the expansion in terms of the Hermite functions presented in [30] (see also [29]). If the expansion functions are supposed generally irreducible to polynomials (the reducible Hermite functions lead to quasi-polynomial solutions much discussed in the context of quasi-exactly solvability [46,47]), this expansion reads [30]

$$u = \sum_{n=0}^{\infty} c_n H_{n+\gamma-\alpha/\varepsilon} \left( \sqrt{-\varepsilon/2} \, (z + \delta/\varepsilon) \right), \tag{32}$$

where the coefficients $c_n$ obey the three-term recurrence relation

$$R_n c_n + Q_{n-1} c_{n-1} + P_{n-2} \, c_{n-2} = 0 , \tag{33}$$

where $\quad R_n = n(\gamma + n - \alpha/\varepsilon)\sqrt{-2\varepsilon} , \quad Q_n = -q - (\gamma + n)\delta , \quad P_n = (\gamma + n)\sqrt{-\varepsilon/2} . \tag{34}$

This series may terminate thus resulting in closed-form solutions [29,30]. This is achieved if $c_{N+1} = c_{N+2} = 0$ (while $c_N \neq 0$) for some $N = 0,1,2,\ldots$. If so, from recurrence relation (33) we have

$$n = N+1: \ R_{N+1} \cdot 0 + Q_N c_N + P_{N-1} \, c_{N-1} = 0 , \tag{35}$$

$$n = N+2: \ R_{N+2} \cdot 0 + Q_{N+1} \cdot 0 + P_N \, c_N = 0 . \tag{36}$$

Thus, it should necessarily be $P_N = 0$ and $Q_N c_N + P_{N-1} c_{N-1} = 0$. The first of these equations gives $\gamma = -N$, while the second one gives a polynomial equation of the degree $N+1$ for the accessory parameter $q$.

For $\gamma = -N$, from the first equation (29) we get $\alpha_0 = (1-N)/2$ and from the first equation (31) we have

$$V_4 = \frac{(N-1)(N+3)\hbar^2}{32m}, \quad N = 0, 1, 2, \ldots. \tag{37}$$

Thus, for all possible sub-potentials of the Lemieux-Bose potential (25) allowing finite-sum Hermite-function solutions the centrifugal term is inverse proportional to $32m$. Another observation is that the only case when this term vanishes is achieved for $N = 1$. Hence, the only case when one may hope to get an *exactly* integrable potential for which no parameter is fixed to a specific value and all involved potential parameters are varied independently, is this case. It is checked that such a particular potential indeed exists. This is the inverse square root potential [34]. Furthermore, it is checked that for the other four Lemieux-Bose potentials [22] this is not the case, that is, they do not possess exactly solvable sub-potentials (for which the solution is achieved by truncation of the Hermite-function expansions) other than the classical harmonic (including the linear sub-case), Coulomb and Morse potentials. The inverse square root potential is thus a rather remarkable exception.

Consider now some particular cases in detail. For $N = 0$ (i.e., when expansion (32) involves just a single term: $u = c_0 H_{\gamma - \alpha/\varepsilon}$) we have the conditions $\gamma = 0$ and $q = 0$. Hence, $V_4 = -3\hbar^2/(32m)$ and from the second equation (30) we get $V_3 = 0$. Thus, the sub-potential of the Lemieux-Bose family (25) allowing a single-term Hermite-function solution is

$$V_{S2} = V_0 + \frac{V_1}{\sqrt{x}} + \frac{V_2}{x} - \frac{3\hbar^2}{32m\,x^2}. \tag{38}$$

This is the second confluent hypergeometric potential by Stillinger [9], Eq.(2.20), which was discussed by several authors in the context of conditionally solvability (see, e.g., [12,48-50]). It has been shown that the "naive" energy spectrum derived through polynomial reductions [48] do not provide physically acceptable wave functions because these solutions do not vanish in the origin [49]. The proper equation for correct eigen-energies is a transcendental equation involving a parabolic cylinder function [9,49].

For $N = 1$, that is, when the expansion (32) is terminated on the second term:

$$u = c_0 H_{\gamma - \alpha/\varepsilon} + c_1 H_{1+\gamma - \alpha/\varepsilon}, \tag{39}$$

the necessary conditions read [30]

$$\gamma = -1, \quad q^2 - \delta q + \alpha = 0. \tag{40}$$

We recall that this is the only case when the centrifugal term vanishes: $V_4 = 0$. Furthermore, the $q$-equation (40) results in $V_2 = 8mV_3^2 / \hbar^2$, so that we obtain the potential

$$V_{E1} = V_0 + \frac{V_1}{\sqrt{x}} + \frac{8mV_3^2 / \hbar^2}{x} + \frac{V_3}{x^{3/2}}. \tag{41}$$

This is the first Exton potential [42], Eq. (21), which has been recently treated in [39]. As particular cases, this potential involves the inverse square root potential [34] (reproduced if $V_3$ is put zero) and the conditionally solvable potential of López-Ortega [37] (achieved by putting $V_1 = 0$). To fulfill the development for $N = 1$, we note that the expansion coefficient $c_1$ of the truncated solution (39) is readily found from the recurrence (33),(34) to be

$$c_1 = -\frac{P_0}{Q_1} c_0 = \frac{\sqrt{-\varepsilon / 2}}{q} c_0. \tag{42}$$

Consider now the cases with $N \geq 2$. To give an explicit representation, here are the corresponding conditions for $N = 2, 3, 4$:

$N = 2$: $\gamma = -2$,

$$q^3 - 3\delta q^2 + 2(\delta^2 + \varepsilon + 2\alpha)q - 4\alpha\delta = 0, \tag{43}$$

$N = 3$: $\gamma = -3$,

$$q^4 - 6\delta q^3 + \left(10\alpha + 11\delta^2 + 10\varepsilon\right)q^2 - 6\delta\left(5\alpha + \delta^2 + 3\varepsilon\right)q + 9\alpha\left(\alpha + 2\left(\delta^2 + \varepsilon\right)\right) = 0, \tag{44}$$

$N = 4$: $\gamma = -4$,

$$\begin{aligned}
&q^5 - 10\delta q^4 + 5\left(4\alpha + 7\delta^2 + 6\varepsilon\right)q^3 - 2\delta\left(60\alpha + 25\delta^2 + 69\varepsilon\right)q^2 - \\
&32\alpha\delta\left(4\alpha + 3\delta^2 + 9\varepsilon\right) + 8\left(8\alpha^2 + 26\alpha\delta^2 + 3\delta^4 + 24\alpha\varepsilon + 18\delta^2\varepsilon + 9\varepsilon^2\right)q = 0.
\end{aligned} \tag{45}$$

As it is seen, the equations become more and more complicated. Checking now if these equations are satisfied by parameters (29)-(31), we reveal that $N = 2$ produces a transformed version of the Exton potential, while the $q$-equation (44) for the case $N = 3$ is not satisfied at all. Next comes $N = 4$ which produces the potential (2) under consideration. Before passing to the details for this case, it should be noted that we have checked that the cases $N = 5$ and $N = 6$ produce rather restrictive potentials which involve all four power-terms of the Lemieux-Bose potential (25) with strengths that depend on a single parameter. Furthermore, the next two cases $N = 7$ and $N = 8$, perhaps unexpectedly, do not produce energy-independent potentials. It is not clear if there are potentials for $N \geq 9$. The $q$-equations become too complicated for checking even with the aid of computer algebra systems.

Besides, there is not a general expression for the $q$-equation because this equation is derived as a determinant of a $N \times N$ matrix involving the coefficients $c_n$ of the expansion (32). Since these coefficients obey a three-term recurrence relation, their explicit forms are not known. Hence, there is a little hope to give a general treatment for the case of larger $N$. These observations explain why we discuss the potential (2). This is the only simple member of the Lemieux-Bose potential (25) (apart from the Stillinger and Exton potentials which have already been studied) allowing a tractable solution. We recall that this potential corresponds to $N = 4$, hence, it is solved by a five-term Hermite-function expansion.

A few complementary details concerning this case are as follows. For $N = 4$ we have $V_4 = 21\hbar^2 / (32m)$. Furthermore, the $q$-equation (45) is shown to take the form

$$V_3 \left( E + f(V_{0,1,2,3}) \right) - \frac{2}{3} V_2 V_1 = 0 \tag{46}$$

with a polynomial function $f$ which does not depend on energy. Requiring now the potential to be energy-independent, we get $V_3 = 0$ and $V_2 V_1 = 0$. The case $V_1 = 0$ is not interesting because it gives a Coulomb plus centrifugal-barrier potential (we note that in this exceptional case, owing to the particular value of the strength of the centrifugal term, the solution is eventually reduced to a single Hermite function). Hence, we take $V_2 = 0$. This recovers the potential (2) under consideration.

The solution of the Schrödinger equation is written as a linear combination of five Hermite functions:

$$\psi_F(z) = z^{-2/3} e^{\alpha_1 z + \alpha_2 z^2} \sum_{n=0}^{4} c_n H_{n+\gamma-\alpha/\varepsilon} \left( \sqrt{-\varepsilon/2} \, (z + \delta/\varepsilon) \right) \tag{47}$$

with arbitrary $c_0$, and $c_{1,2,3,4}$ being calculated using the recurrence (33),(34). The last step is to apply the recurrence relation

$$H_{\nu+1}(z) = 2zH_\nu(z) - 2\nu H_{\nu-1}(z) \tag{48}$$

to reduce the linear combination of the five Hermite functions to a combination involving only two such functions. With a simplification for the case when $V_1 < 0$ and $E < V_0$, this recovers the fundamental solution (3),(4) with parameters (5),(6).

## 5. Discussion

Thus, we have introduced a bi-confluent Heun potential conditionally integrable in terms of the confluent hypergeometric functions. We have presented the exact solution of the Schrödinger equation for this potential written as a linear combination with non-constant coefficients of two non-integer order Hermite functions. Discussing the bound states supported by the potential, we have derived the exact equation for the energy spectrum. We have suggested a modified approximation for the Hermite functions involved in the spectrum equation. The modification, which provides a highly accurate result, involves two variable constants which are determined by matching the exact Hermite function and the approximation in an asymptote and in a finite point close to the origin. Using this result, we have approximated the spectrum equation by a transcendental equation involving trigonometric functions. Finally, we have constructed a highly accurate approximation for the energy levels. We have seen that the bound-state wave-functions are not (quasi) polynomials because the indexes of the involved Hermite functions are not integers.

A peculiarity of the presented potential worth to be discussed a bit more is that the solution of the Schrödinger equation is not written through a single-term ansatz involving just one hypergeometric function. In fact, the class of potentials suggested by irreducible multi-term solutions is much richer as compared to the class provided by the single-term ansatz. A simplest possibility suggesting an infinite set of potentials having a multi-term solution is achieved by discussing piecewise continuous potentials involving fragments each of which belongs to a separate potential which is itself solved through a single-term ansatz [51]. Since the general solution for each separate region nevertheless involves two independent fundamental single-term solutions, it is understood that the matching conditions for the boundaries of the intervals on which the separate solutions are defined will lead to solutions that in general are multi-term [51] (see [51-53] for several examples). The solution that we have presented eventually involves two terms. This is because the starting five-term solution involves only contiguous Hermite functions. We conclude by noting that owing to the contiguous function relations this two-term structure is a common feature for all Heun potentials [15,22-25] if the Schrödinger equation is solved in terms of the hypergeometric functions on the original definition interval.

### Acknowledgments

This research has been conducted within the scope of the International Associated Laboratory IRMAS (CNRS-France & SCS-Armenia). The work has been supported by the Armenian State Committee of Science (SCS Grant No. 15T-1C323), Armenian National Science and Education Fund (ANSEF Grant No. PS-4986), the Ministry of Education and Science of the Russian Federation (State Assignment No. 3.873.2017/4.6), and the project

"Leading Russian Research Universities" (Grant No. FTI_24_2016 of the Tomsk Polytechnic University). T.A. Ishkhanyan acknowledges the support from SPIE through a 2017 Optics and Photonics Education Scholarship and thanks the French Embassy in Armenia for a doctoral grant.